# Electronic structure of a graphene-like artificial crystal of NdNiO$_3$

Arian Arab[1], Xiaoran Liu[2], O. Köksal[3], W. Yang[1], R. U. Chandrasena[1], S. Middey[4], M. Kareev[2], S. Kumar[4], M.-A. Husanu[5,6], Z. Yang[7], L. Gu[7,8], V. N. Strocov[5], T.-L. Lee[9], J. Minár[10], R. Pentcheva[3], J. Chakhalian[2], and A. X. Gray[1,*]

[1] *Physics Department, Temple University, Philadelphia, Pennsylvania 19122, USA*

[2] *Department of Physics and Astronomy, Rutgers University, Piscataway, New Jersey 08854, USA*

[3] *Department of Physics and Center for Nanointegration Duisburg-Essen (CENIDE), University of Duisburg-Essen, Duisburg 47057, Germany*

[4] *Department of Physics, Indian Institute of Science, Bengaluru 560 012, India*

[5] *Swiss Light Source, Paul Scherrer Institute, Villigen, Switzerland*

[6] *National Institute of Materials Physics, 077125 Atomistilor 405A, Magurele, Romania*

[7] *Beijing National Laboratory for Condensed-Matter Physics and Institute of Physics, Chinese Academy of Sciences, Beijing 100190, People's Republic of China*

[8] *Collaborative Innovation Center of Quantum Matter, Beijing 100190, People's Republic of China*

[9] *Diamond Light Source Ltd., Didcot, Oxfordshire OX11 0DE, United Kingdom*

[10] *New Technologies-Research Center, University of West Bohemia, CZ-30614 Pilsen, Czech Republic*

*email: axgray@temple.edu

## Abstract

Artificial complex-oxide heterostructures containing ultrathin buried layers grown along the pseudocubic [111] direction have been predicted to host a plethora of exotic quantum states arising from the graphene-like lattice geometry and the interplay between strong electronic correlations and band topology. To date, however, electronic-structural investigations of such atomic layers remain an immense challenge due to the shortcomings of conventional surface-sensitive probes, with typical information depths of a few Ångstroms. Here, we use a combination of bulk-sensitive soft x-ray angle-resolved photoelectron spectroscopy (SX-ARPES), hard x-ray photoelectron spectroscopy (HAXPES) and state-of-the-art first-principles calculations to demonstrate a direct and robust method for extracting momentum-resolved and angle-integrated valence-band electronic structure of an ultrathin buckled graphene-like layer of $NdNiO_3$ confined between two 4-unit cell-thick layers of insulating $LaAlO_3$. The momentum-resolved dispersion of the buried Ni $d$ states near the Fermi level obtained via SX-ARPES is in excellent agreement with the first-principles calculations and establishes the realization of an antiferro-orbital order in this artificial lattice. The HAXPES measurements reveal the presence of a valence-band (VB) bandgap of 265 meV. Our findings open a promising avenue for designing and investigating quantum states of matter with exotic order and topology in a few buried layers.

A central theme threading through the modern condensed matter is the search and understanding of different quantum phases of matter. After decades of intense studies, strong electron correlations, initially neglected in the band theory, have been recognized as the key factor that can drive a system into diverse many-body phases such as unconventional magnetic and superconducting states, metal-to-insulator transition, quantum criticality, and glass-like spin and charge states to name a few.[1] Subsequently, during the past decade, it was realized that strong spin-orbit coupling (SOC), once added to the band structure of a weakly-correlated material, can dramatically influence the electronic states and lead to qualitatively different phases of matter, including topological insulators, topological superconductors, and topological semimetals.[2-5] Inspired by these results, very recently, it has been proposed that merging electronic correlations and SOC can potentially bring about a plethora of exotic quantum states such as axion insulator, topological Mott insulator, multipolar order, Kitaev spin liquid, and so on.[6-10] Furthermore, search for the materials with singularities in the density of states stemming from internal lattice symmetries and fine-tuned interactions have attracted a particular interest as the possible platform for topologically non-trivial interacting states of matter, including quantum anomalous Hall (QAH) effect and topological superconductivity.[11-13]

From the experimental standpoint, however, the realization of the correlated topological states runs into a major challenge of finding bulk crystals to closely match the theoretical proposals. With the recent advances in synthesis of ultra-thin layers of chemically-complex materials, this challenging issue could be potentially mitigated. Specifically, the fabrication of a specific number of unit cells of an ultra-thin film derived from perovskite or pyrochlore structure along high-index crystallographic directions naturally gives rise to the buckled graphene-like or frustrated lattice geometry, and results in interactions favoring the correlated topological phases.[7,10,14-16] To date, however, two key obstacles stand in the path of realization of such a design paradigm: (1) the ongoing difficulties of the growth along polar directions[17-20] and (2) the critical problem of measurement of electronic band structure from only one or two unit cells buried under a non-active capping layer. As a result, the majority of conventional surface-sensitive probes such as angle-resolved photoemission spectroscopy (ARPES), which provides direct information on the electronic bands, are severely limited in their applicability for such material systems. Once resolved, the exquisite control over lattice geometry and electronic structure stemming from the successful resolution of these two challenges will enable the next generation of artificial topological materials with strong electronic correlations.

In this letter, we use a combination of bulk-sensitive soft x-ray angle-resolved photoelectron spectroscopy (SX-ARPES),[21,22] hard x-ray photoelectron spectroscopy (HAXPES)[23] and density functional theory (DFT+$U$) calculations to investigate the momentum-resolved and angle-integrated valence-band electronic structure of the buried two-dimensional $NdNiO_3$ layer in a geometrically-engineered superlattice grown epitaxially on a $LaAlO_3$ (111) oriented single-crystal substrate.[24] In this system, the nickelate layer forms a buckled honeycomb lattice (see Fig. 1a) that was initially proposed to give rise to a characteristic band structure consisting of four bands around the Fermi level: two linearly crossing at K forming a Dirac point, and two flat with a quadratic touching point of the two sets at Γ for ferromagnetic (FM) coupling.[25-27] More recent DFT+$U$ calculations showed that a bond disproportionation in a form of breathing-mode distortion of the oxygen octahedra lifts the equivalence of the two Ni sublattices and opens a small gap at the Dirac (K, K') points, thus destroying the topological QAH/FM phase and shifting the material toward the Mott part of the phase diagram.[28,29] Furthermore, x-ray resonant magnetic scattering measurements found indication for antiferromagnetic (AFM) spin correlations instead.

Polarization-dependent x-ray absorption spectroscopy (XAS) measurements[24] revealed almost an order of magnitude stronger x-ray linear dichroism (XLD) effect than that observed in (001)-oriented superlattices.[30-32] As suggested by the first-principle calculations, such surprising observation is related to the appearance of an antiferro-orbital ordering of $d_{3z^2-r^2}$ and $d_{3x^2-r^2}$ orbitals, stemming from the lowering of structural symmetry to P1.[24] This unusual effect originates from the decoupling of the two Ni sublattices due to AFM order, where the decoupling of the $NiO_6$ octahedra is realized by the surrounding $AlO_6$ octahedra in the double perovskite structure. Theoretically, two different orbital arrangements are possible: one with P1 symmetry in a 1×1 unit cell and another with P3 symmetry in a larger √3×√3R30° lateral unit cell (see Fig. S1 in the Supporting Information). Here, we provide direct unambiguous evidence by the SX-ARPES measurements supporting the P1 symmetry, which is consistent with the results suggested by the XLD data.

For our experiments, high-quality epitaxial superlattice consisting of three repetitions of (111)-oriented [2 u.c. $NdNiO_3$ / 4 u.c. $LaAlO_3$] (with $LaAlO_3$ layer on top) was grown by pulsed laser interval deposition[33] on a $LaAlO_3$ (111) substrate (Fig. 1a). Coherent epitaxy and thicknesses of all layers were monitored during growth using reflection high-energy electron diffraction (RHEED) and confirmed *ex-situ* using scanning transmission electron microscopy (STEM) (Fig. 1b). The quality (flatness) of the surface was verified using atomic force microscopy (see Fig. S2 in the Supporting Information). Crystallinity and correct layering of the superlattice was confirmed using high-resolution synchrotron-based x-ray diffraction spectroscopy (see Fig. S3 in the Supporting Information). Additional details regarding synthesis and characterization are provided in the Methods section of the Supporting Information with relevant references to prior studies focusing on optimization and characterization of this and similar [111] multilayer structures.

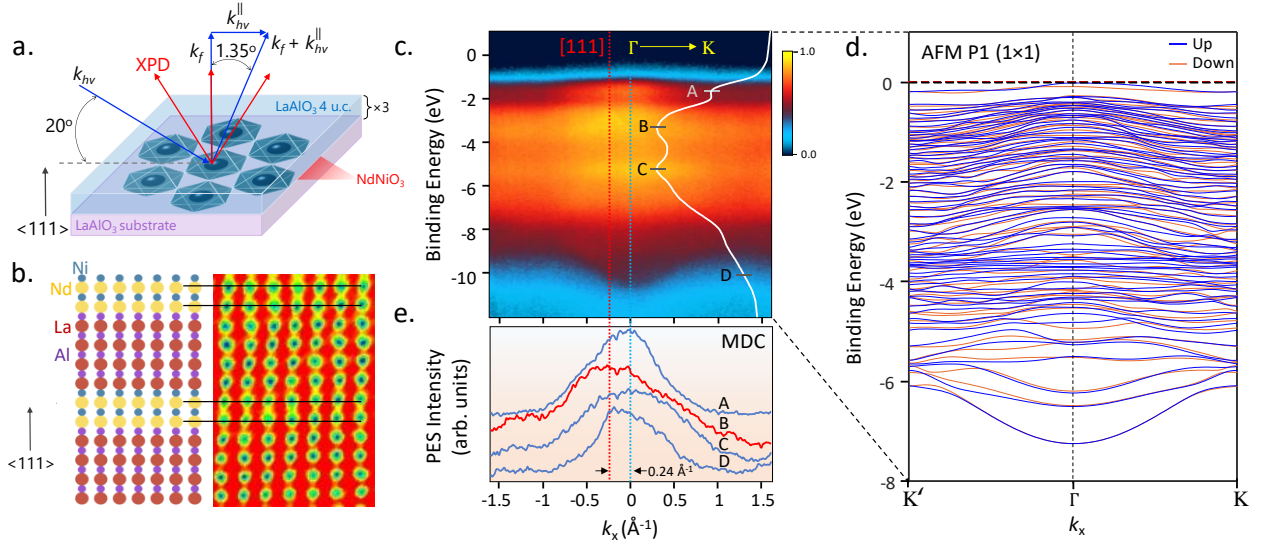

**Figure 1.** SX-ARPES spectroscopy and band-structure calculations. (a) Schematic diagram of the SX-ARPES experimental geometry with all relevant momentum vectors shown and labeled. (b) High-resolution STEM cross sectional image of the sample and the corresponding schematics. (c) SX-APRES spectrum of the valence-band dispersion measured along the K'-Γ-K high-symmetry direction. The white EDC curve represents the momentum-integrated MEW-DOS, with most prominent features A-D labeled. (d) The DFT+$U$ calculation of the spin-projected band structure for the AFM P1 (1×1) configuration shown for the same cut of the BZ. **e.** MDC curves recorded at the binding energies of the above-mentioned features.

Soft x-ray angle-resolved photoemission spectroscopy measurements were carried out at the high-resolution ADRESS beamline of the Swiss Light Source (see also Methods section of the Supporting Information and Ref. 34 for details). By carrying out the measurements at the photon energies between 642 and 874 eV we effectively increase the inelastic mean-free paths (IMFP) of the photoelectrons inside the superlattice by a factor of 3-5 compared to typical ARPES studies ($hv$ = 30-100 eV),[35] significantly enhancing the information depth and thus permitting momentum-resolved study of the artificial quantum layer of NdNiO$_3$ (111) sandwiched between two slabs of insulating LaAlO$_3$ (111). All the measurements were done in the near-normal emission (NE) experimental geometry depicted schematically in Fig. 1a, with the x-ray grazing incidence angle of 20° as measured from the sample surface. Fig. 1c shows a valence-band dispersion along the K'-Γ-K high-symmetry direction recorded at the photon energy of 642 eV, placing the final photoelectron wave vector $\mathbf{k}_f$ close to the Γ point along the $k_z$ direction in the extended BZ picture. This is also confirmed experimentally with a variable photon energy ($k_z$) scan (see Fig. S4 in the Supporting Information). The spectra in Fig. 1c are strongly modulated by the photoemission matrix elements in energy, resulting in the matrix-element-weighted density of states (MEW-DOS), which is represented by a white energy distribution curve (EDC) on the right side of the plot, with the four most prominent features appearing at -1.6, -3.3, -5.1 and -10.0 eV, and labeled A through D, respectively. The experimental momentum dispersion exhibits excellent agreement with the band structure calculated with DFT+$U$ for the AFM ordering of the Ni sites with the P1 symmetry (Fig. 1d). The valence-band dispersion is dominated by two flat bands just below the Fermi level (one in each spin channel), a dense set of relatively flat bands in the intermediate region between -0.3 and -5 eV, followed by several more dispersive bands at the bottom of the valence band with a minimum at Γ, corresponding to feature D in Fig. 1c. The total width of the calculated valence-band manifold is slightly underestimated, as evidenced by the different binding energy scale - a general feature of this calculation scheme, wherein parameters such as the bandgap, $d$-band bandwidths and positions are very sensitive to the changes in the Hubbard terms (see *e.g.* a systematic study of these effects in ZnO in Ref. 36). Additional comparison of the experimental data with the theoretical band-structure calculations, contrasting the results of the latter for the P1 and P3 symmetries, are shown and discussed in the Supporting Information (see Fig. S5).

The momentum distribution curves (MDC) recorded at the binding energies of the four above-mentioned features A-D (Fig. 1e) reveal an important effect, which is characteristic of the higher-energy (soft- and hard x-ray) ARPES and is depicted schematically in Fig. 1a.[37,38]

Specifically, as the dipole approximation, which is typically used in ARPES, no longer fully holds for the higher photon energies, the non-negligible photon momentum wave vector ($\mathbf{k}_{hv}$) shifts the position of the final photoelectron wave vector ($\mathbf{k}_f$) along the $k_x$ direction in accordance with a simple wave vector conservation equation $\mathbf{k}_f = \mathbf{k}_i + \mathbf{k}_{hv}^{\parallel}$, where $\mathbf{k}_i$ is the initial-state wave vector and $\mathbf{k}_{hv}^{\parallel} = \mathbf{k}_{hv} \cos(20°)$ is the photon momentum wave vector adjusted for the grazing incidence angle of 20°. As a result, electronic dispersion observed in the 2D detector is shifted relative to the normal [111] direction (red dashed line) by approximately 0.24 Å$^{-1}$, in excellent agreement with the theoretically predicted shift of 0.3 Å$^{-1}$ (or 1.35° in the detector-angle units). This shift is immediately apparent to the eye for feature D, due to its strongly-dispersive nature. From Fig. 1c, it is evident that this parabolic band is clearly centered around the $\mathbf{k}_{hv}$-shifted Γ point (blue dashed line). Data in Fig. 1c can be normalized by the binding-energy-averaged and $k$-averaged spectra to enhance the dispersive features, clearly bringing out the parabolic character of feature D (see Fig. S5b of the Supporting Information).

In stark contrast with the strongly-dispersive feature D, the intensity of a mostly non-dispersive feature B is centered around the normal [111] direction (red dashed line in Figs. 1c and e). The angular photoelectron intensity distributions for such mostly-flat features are dominated by the x-ray photoelectron diffraction (XPD) effects and not affected by the photon momentum ($\mathbf{k}_{hv}$) shift, analogous to localized core-levels with no dispersion in $k$.[39]

As an intermediate case, feature A, which is weakly dispersive (see Fig. 1c), shows a mixture of XPD and electronic band structure characteristics, as evident from its MDC, which is centered in-between the [111] and Γ directions, with its maximum somewhat closer to the Γ point. For such features, it is critical to emphasize the importance of utilizing data normalization procedures that separate the effects of true electronic-state dispersion from XPD in soft- and hard x-ray ARPES (Figs. S5a-b and S7 of the Supporting Information, as well as Refs. 38 and 40).

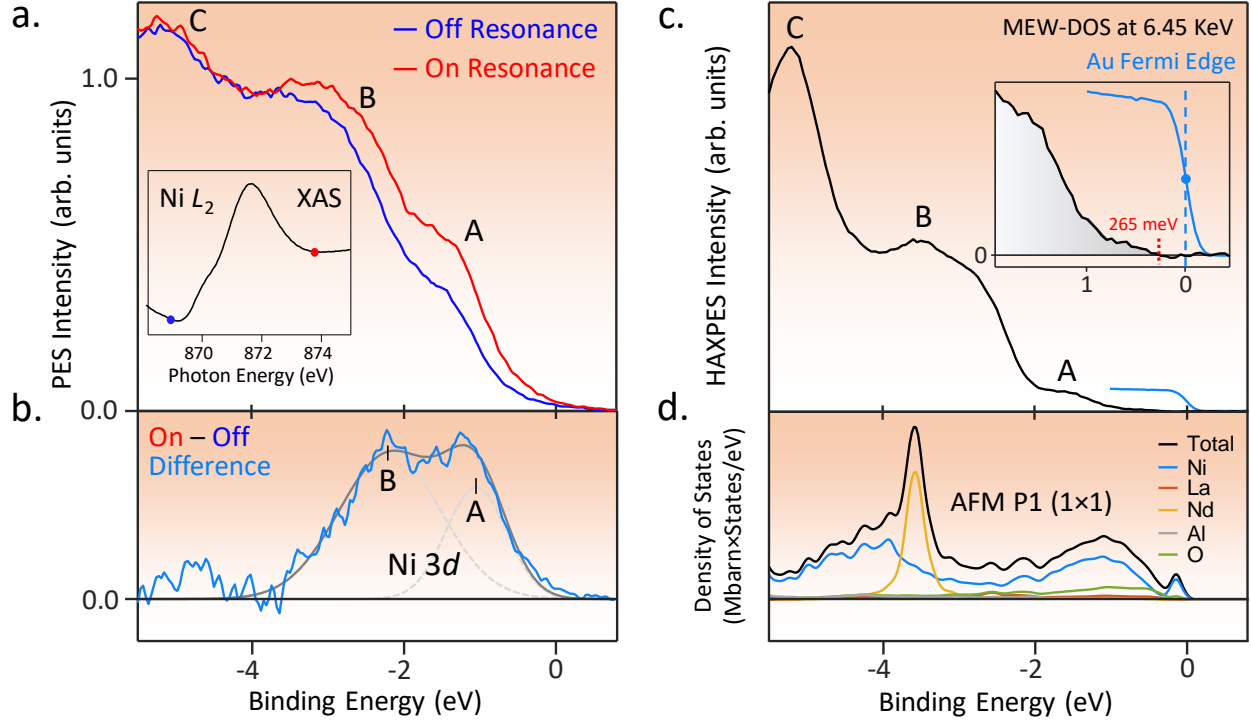

**Figure 2.** Resonant (Ni $L_2$) valence-band spectroscopy and HAXPES. (a) Angle-integrated VB spectra recorded at the photon energies of 868.5 eV (off resonance) and 873.4 eV (on resonance) reveal the contribution of the Ni 3$d$ states via resonant enhancement. Inset shows the Ni $L_2$ XAS spectrum with the blue and red markers at the relevant photon energies. (b) The difference spectrum obtained by subtracting the 'off' spectrum shown in panel **a.** from the 'on' spectrum. Two distinct components A and B, corresponding to the NdNiO$_3$-derived Ni 3$d$ states, are fitted using two Gaussian peaks centered at -1.1 and -2.2 eV. (c) Bulk-sensitive HAXPES spectrum recorded at the photon energy of 6.45 keV with the IMFP of approximately 85 Å.[35] Inset shows a high-statistics spectrum of the valence-band maximum (at -265 meV), referenced to the Au Fermi edge. (d) Cross-section-weighted element-projected and total DOS of the superlattice, calculated in the GGA+$U$ framework of DFT, and broadened by convolution with Gaussian and Lorentzian functions to account for both experimental and hole lifetime broadening.

Fig. 2a shows the VB photoemission spectra recorded at the photon energies of 868.5 eV and 873.4 eV, corresponding to the 'off' and 'on' resonant conditions for the Ni $L_2$ absorption edge, respectively. The corresponding XAS spectrum, measured *in-situ* using the total electron yield detection mode, is shown in the inset. Resonant photoemission at the transition-metal $L_{2,3}$ edges takes advantage of the interference between the direct photoemission channel $2p^63d^n \rightarrow 2p^63d^{n-1} + e^-$ and the decay of a resonantly-excited state $2p^53d^{n+1} \rightarrow 2p^63d^{n-1} + e^-$, which leads to the enhancement of the 3d photoemission cross section.[41] As a result, in our case the contribution from the Ni 3d states in the valence bands is significantly amplified, as evident from the "on resonance – off resonance" difference spectrum shown in Fig. 2b. The above-mentioned difference spectrum exhibits two distinct resonant features labeled A and B near the Fermi level (at -1.1 and -2.2 eV), corresponding to the strongly-hybridized Ni 3d $e_g$ and $t_{2g}$ states, as well as some additional resonant enhancement of photoemission intensity between -4 and -5 eV, fully-consistent with the first-principles DFT+*U* calculations shown in Fig. 2d and discussed below. We note that the Ni $L_2$ resonance had to be used instead of $L_3$, due to the strong overlap of the latter with the La $M_4$ absorption edge.[42]

In order to probe the entire depth of the sample, we carried out complementary angle-integrated HAXPES measurements of the valence bands at a photon energy of 6.45 keV, at which the IMFP is estimated to be approximately 85 Å,[35] thus allowing to directly probe the MEW-DOS of the entire superlattice and facilitating a straightforward comparison of the experimental data to theory (see Methods section of the Supporting Information for details). The kinetic energy of the Fermi level was determined in-situ with a high-statistics measurement on a clean sputtered thin-film Au sample, thus allowing for an accurate calibration of the binding energies for both HAXPES and SX-ARPES measurements.

The experimental HAXPES VB spectrum, shown in Fig. 2c exhibits excellent agreement with the total DOS obtained using first-principles DFT+*U* calculations (black spectrum in Fig. 2d) both in terms of the relative intensities and relative positions of the key features. Furthermore, the experimental valence-band maximum appears at the binding energy of -265 meV ($E_{\text{VB-max}}$) below the Fermi level, as referenced by the Au Fermi-edge measurement (inset). The magnitude of this value corresponds to the size of the valence-band bandgap ($E_{\text{VB-max}} - E_F$), suggesting that the full bandgap of NdNiO$_3$ (111) is at least this large. Indeed, the DFT+$U$ calculations render a full bandgap (between the valence-band maximum and the conduction-band minimum) of approximately 1 eV (see Fig. S6 in the Supporting Information). Finally, the photoionization-cross-section-weighted[43] and element-projected DOS spectra in Fig. 2d (color spectra) reveal the dominant character of the NdNiO$_3$-derived states within the valence bands. In particular, while the Nd *4f* states contribute to a strong localized peak at -4 eV, the Ni *3d* states hybridized with oxygen dominate the whole width of the valence bands with $e_g$ character just below the Fermi level and extending up to the bottom of the VB (not shown here), and Ni $t_{2g}$ states in between, in agreement with the soft x-ray resonant photoemission results in Figs. 2a-b.

Figure 3a shows the XPD-corrected[38,40] isoenergetic SX-ARPES intensity map in ($k_x$, $k_y$) integrated over a 200 meV binding-energy window, which is comparable with our total experimental energy resolution of 111 meV, and centered at the binding energy of the VB feature A (Ni *3d* peak at -1.6 eV). The map was obtained by recording the $k_x$-$E_b$ dispersions while rotating the sample about the polar 'tilt' axis, which is orthogonal to $k_x$ and thus corresponds to the $k_y$ direction in the momentum space.[40] To access a wider range of $k_x$ values, angle-resolved measurements were repeated at the photoelectron take-off angles of +/-4° from the normal [111] direction. The data was combined with the NE measurement to form a larger, more complete dataset. Faint vertical lines at $k_x$ = +/-2 Å$^{-1}$ in Fig. 3a mark the places where the datasets overlapped. The XPD correction, resulting in the separation of the electronic band dispersion from the residual

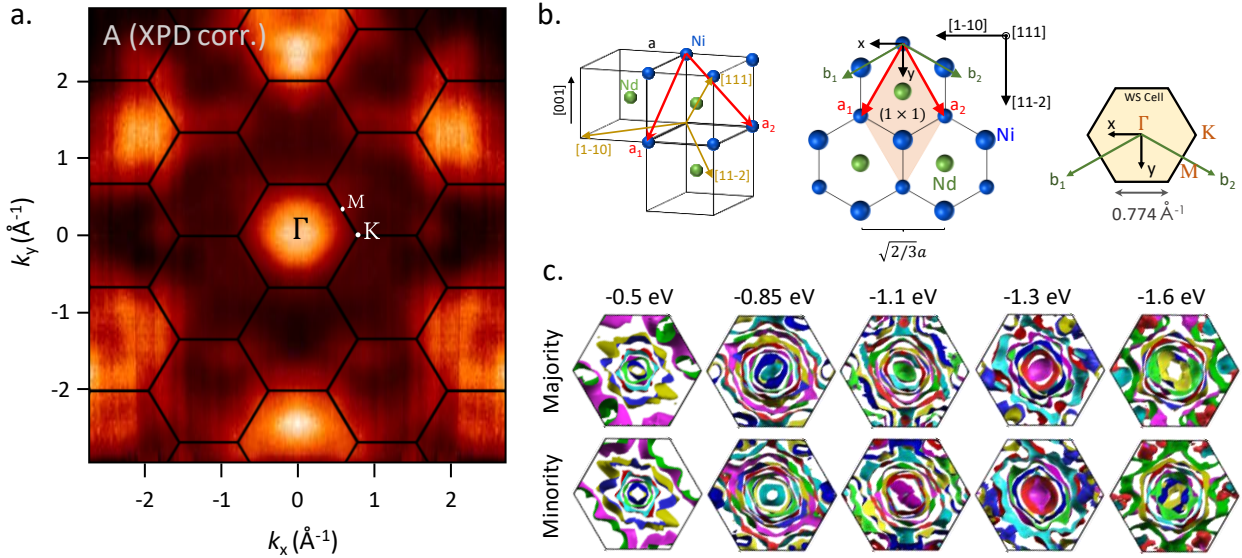

**Figure 3.** Isoenergetic $k_x$-$k_y$ maps: SX-ARPES experiment and theory. (a) Momentum-resolved XPD-corrected SX-ARPES photoemission intensity map of the Ni 3$d$ states near the valence-band maximum (feature A, centered at -1.6 eV below $E_F$). (b) Perovskite structure along conventional [001] direction on the left and along [111] direction in the middle are shown; right panel shows the hexagonal Brillouin Zone of the [111] structure. (c) Isoenergetic cuts through the band structure in reciprocal space for the majority (top row) and minority (bottom row) bands, calculated for the binding-energy range from -0.5 eV to -1.6 eV, and spanning the major Ni 3$d$ features near the VB maximum.

XPD intensity modulations, was carried out according to the procedure described in the Supporting Information (see Fig. S7) and consistent with that used in prior studies.[38,40]

The calculated schematic map of the extended BZ picture zone boundaries, with the central BZ at (0 Å$^{-1}$, 0 Å$^{-1}$), is overlaid on the experimental data, matching perfectly the hexagonal symmetry of the observed photoemission intensity distribution, which is consistent with the buckled honeycomb NdNiO$_3$ (111) structure (see Fig. 3b). Minor apparent distortions in the image are observed near the edges of the map, since the final photoelectron wave vector spans a spherical (and not planar) trajectory in the three-dimensional extended BZ picture.[38] Additionally, the neighboring BZs alternate in the overall photoemission intensities due to the matrix-element effects, which are ubiquitous in angle-resolved photoemission experiments.[44,45] This inherent experimental artefact, leading to suppression of the photoemission intensities for every other Brillouin zone along the Γ-M high symmetry directions, has been recently observed in similar [111] compounds,[46,47] as well as other solids with hexagonal crystalline structure.[48,49]

The observed periodicity and the BZ size correspond to the simulated AFM P1 symmetry, as shown schematically in Fig. 3b, providing additional evidence for the latter. In Fig. 3c, we show isoenergetic cuts through the band structure in reciprocal space for the majority and minority bands, calculated for the energy range spanning the near-$E_F$ Ni 3$d$ states (from -0.5 eV to -1.6 eV). Additional calculations for a wider energy range, from -0.075 eV to -3.58 eV, are shown in Fig. S8 of the Supporting Information. The isoenergetic cuts are dominated by the hexagonal/circular features around Γ and additional features centered at the zone boundary at K. Although some of the finer details of the ($k_x$, $k_y$) band dispersion observed in theory are averaged out in the experiment because of our total energy resolution (~111 meV), we observe a good agreement between the two.

As an additional theoretical benchmarking test, we compared our results to the one-step photoemission calculations based on a fully relativistic LDA+$U$ layer-KKR approach and a time-reversed LEED final state.[50,51] Results of these additional calculations are shown in Fig. S9 of the Supporting Information and represent true angular distributions of photoemission intensities in the extended BZ picture, as applied to the actual multilayer structure with the explicitly included surface. Good similarities between the experiment and the simulation are observed: in particular, the sizes and the periodicity of the BZs that are consistent with the 1×1 P1 structure modeled for our specific experimental geometry, along with the above-mentioned depressions in the intensities of adjacent BZs due to the matrix-element effects. Such calculations represent a key first step in the interpretation of the bulk-sensitive momentum-resolved photoemission data for [111] oriented multilayers.

## ACKNOWLEDGEMENTS

A.X.G., A.A., W.Y. and R.U.C. acknowledge support from the U.S. Army Research Office, under Grant No. W911NF-15-1-0181. A.X.G. also acknowledges support from the US Department of Energy, Office of Science, Office of Basic Energy Sciences, Materials Sciences and Engineering Division under award number DE-SC0019297 during the writing of this paper. X.L and J.C. are supported by the Gordon and Betty Moore Foundation EPiQS Initiative through Grant No. GBMF4534. R. P. and O.K. acknowledge funding by the German Science Foundation within CRC/TRR80, project G3. S.M. acknowledges support from ISRO-IISc Space Technology Cell during the writing of this paper. J.M. would like to thank CEDAMNF project financed by Ministry of Education, Youth and Sports of Czech Rep., project No. CZ.02.1.01/0.0/0.0/15.003/0000358. The authors thank Phillip Ryan (Argonne National Laboratory) for assistance with the x-ray diffraction measurement.

## Supporting Information: Methods

**Sample Synthesis and Characterization.** [2 u.c. NdNiO$_3$ / 4 u.c. LaAlO$_3$] ×3 superlattice was grown on LaAlO$_3$ (111) in layer-by-layer manner using a pulsed laser interval deposition system, equipped with high-pressure RHEED.[1,2] The growth was carried out at 670°C under 150 mTorr partial pressure of oxygen. The sample was annealed at the growth temperature for 30 minutes under 1 bar pressure of ultra-pure oxygen. The STEM measurements were carried out using a spherical aberration-corrected JEM-ARM200F microscope operated at 200 kV in the high-angle annular dark field (HAADF) imaging mode. The flat morphology of the surface was verified using AFM (see Fig. S2), revealing RMS roughness of approximately 200 pm (less than one unit cell). Crystallinity and correct layering of the superlattice was confirmed using high-resolution synchrotron-based x-ray diffraction (XRD) spectroscopy (see Fig. S3). In addition to this, an exhaustive characterization of this and similar [111] superlattices via in-situ RHEED, STEM, XRD, XAS, and AFM was carried out by us in prior studies, confirming crystallinity, layering, phase purity, coherence, morphology and relevant chemical/valence states of the constituent cations.[1,2,3]

**Bulk-Sensitive Photoemission Spectroscopy.** The SX-ARPES endstation at the high-resolution ADRESS beamline at the Swiss Light Source was equipped with a SPECS PHOIBOS-150 hemispherical electrostatic analyzer and a six-axis cryogenic manipulator, allowing for facile three-dimensional mapping of the valence-band electronic structure in the momentum space ($k_x$, $k_y$ and $k_z$ via variable photon energy). At the photon energy of 642 eV, total instrumental energy resolution was estimated to be 111 meV. The HAXPES measurements were carried out at Beamline I09 of the Diamond Light Source at the photon energy of 6.45 keV and with a comparable total experimental energy resolution of approximately 200 meV. It is important to note that due to the much higher photon energies utilized in SX-ARPES and HAXPES, the effective exprimental resolutions in both energy and momentum are limited, as compared to conventional high-resolution ARPES (~1-5 meV). Our measurements represent the current status of the state-of-the-art for both SX-ARPES[4,5] and HAXPES.[6] Significant efforts in the instrumentation improvements of both the x-ray sources (beamlines/monochromators) and the photoelectron detectors (analyzers) are currently underway at several facilities, indicating that resolutions of <50 meV (for HAXPES) and <30-40 meV (for SX-ARPES) are achievable in the near future.

All the measurements were carried in the near-normal emission (NE) [111] experimental geometry. For the angle-resolved SX-ARPES measurements, the fixed angle of 70° between the beamline and the analyzer direction thus required the grazing incidence angle to be set at 20° (= 90° - 70°). The sample temperature was set to 100K.

**Theoretical Calculations.** Density functional theory (DFT+$U$) calculations were carried out with the all-electron full potential linearized augmented plane wave method in the WIEN2k code.[7,8] The generalized gradient approximation[9] was used for the exchange-correlation functional together with an on-site Hubbard $U$ term[10] with $U$ = 5.0 eV and J = 1.7 eV for Ni 3$d$ and $U$ = 8 eV for Nd. The 1×1 unit cell with P1 symmetry was modelled with a 30 atom supercell with the lateral lattice constant of LaAlO$_3$, while the √3×√3R30° reconstruction required a three-times-larger unit cell. Octahedral tilts and distortions were fully taken into account. The Fermi surface calculations for different isoenergetic levels were carried out with a very dense 30×30×10 $k$-point grid and plotted using the XCRYSDEN package.[11]

The SX-ARPES calculations were carried out within the framework of the fully-relativistic one-step model of photoemission,[12] as implemented in the multiple scattering Green function SPR-KKR package.[13] This method accounts for all relevant experimental aspects of the photoemission process, including experimental geometry, photon energies, x-ray polarization, matrix element effects, multiple scattering in the initial and final states, and all surface- and interface-related effects in the excitation process.

# Figure S1

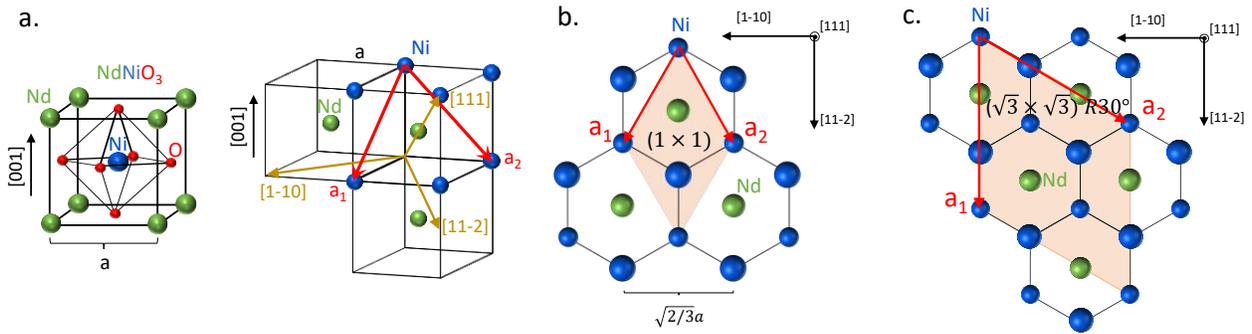

**Figure S1.** Structural diagram depicting the perovskite structure along the conventional [001] direction (a), and the difference between the P1 (1×1) and P3 (√3×√3) symmetries (b and c). For clarity, the structure is depicted in the pseudocubic notation (a) and the oxygen atoms are not shown (b and c). The diagram defines the sizes of the corresponding unit cells. The differences in orbital orientation of the Ni1 and Ni2 $3d$ orbitals which lead to the larger unit cell of the P3 (√3×√3) symmetry are shown in Fig. 4 of Ref. 24 of the main text.

# Figure S2

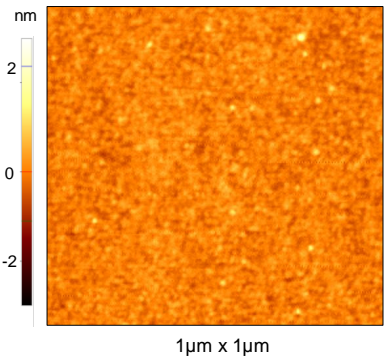

**Figure S2.** Atomic force microscopy (AFM) measurements of the [2 u.c. NdNiO$_3$ / 4 u.c. LaAlO$_3$] (111) superlattice grown on an LaAlO$_3$ (111) substrate. The RMS surface roughness was found to be approximately 200 pm (less than one unit cell), which confirms smooth surface morphology of the structure.

# Figure S3

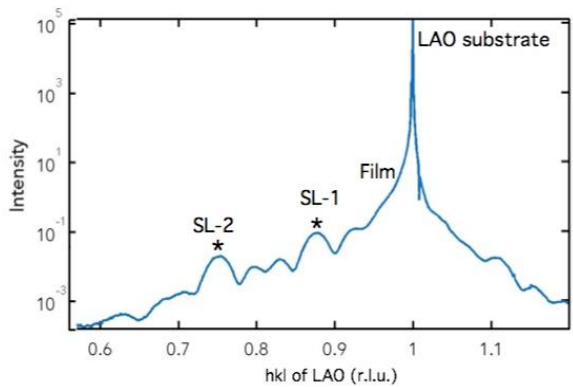

**Figure S3.** High-resolution synchrotron-based XRD spectroscopy measurements confirm growth along the [111] direction and exhibit thickness fringes as well as superlattice reflections (marked by *), as expected for a 2NNO/4LAO repeat unit (6 unit cells). The results suggest correct layering as well as atomically sharp interfaces and absence of chemical disorder over a macroscopically-large length scale.

**Figure S4**

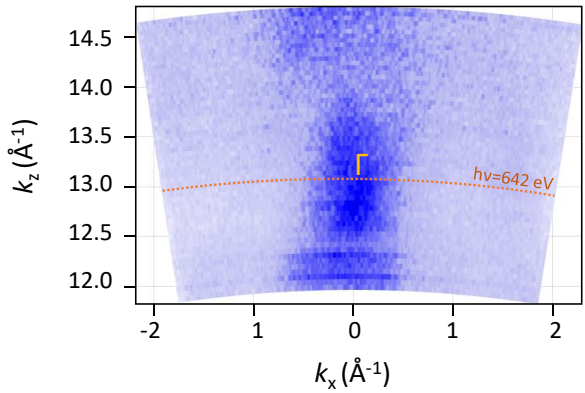

**Figure S4.** Momentum-resolved $k_x$-$k_z$ map obtained by scanning the excitation x-ray photon energy. At the photon energy of 642 eV, the final photoelectron wave vector $k_f$ points close to a high-symmetry point along the $k_z$ direction in the extended BZ picture.

**Figure S5**

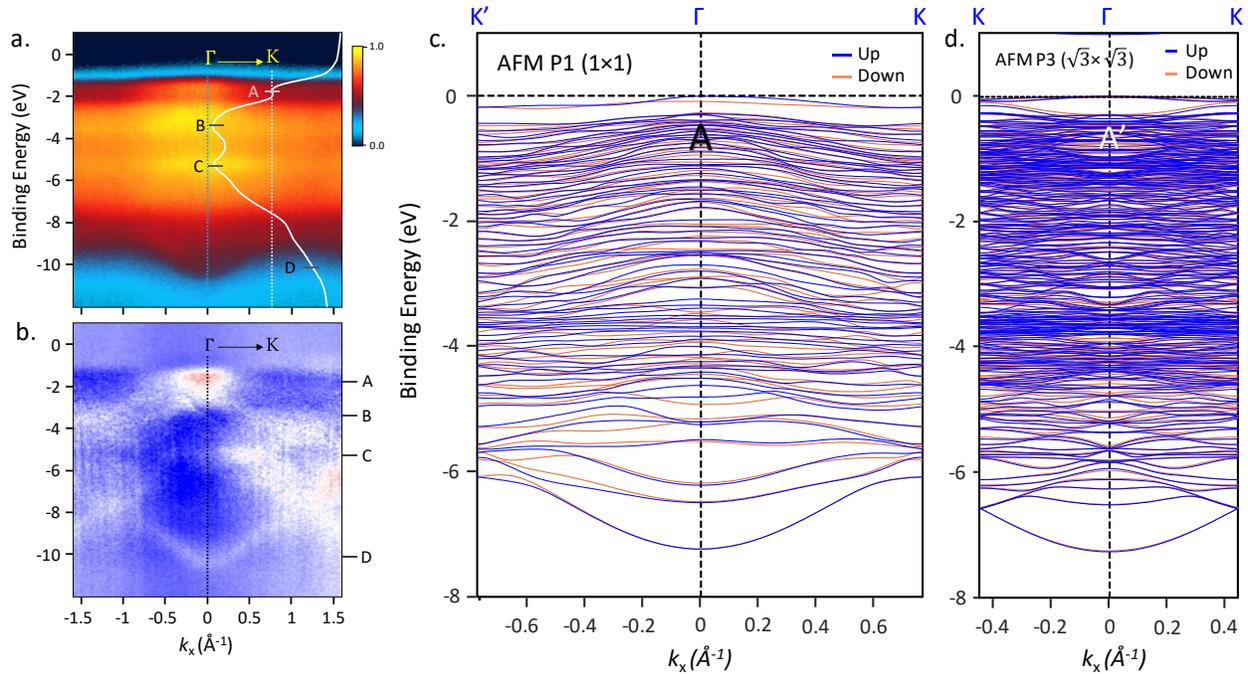

**Figure S5.** Comparison of the experimental data (a and b) and the band structure calculations for the P1 (1×1) and P3 (√3×√3) symmetries (c and d). In panel b, the raw spectrum was normalized by the binding-energy-averaged and the $k$-averaged spectra to enhance the dispersive features (*e.g.* A and D), while suppressing the non-dispersive features (*e.g.* B and C).

**Discussion**

One of the differences allowing to discriminate between the P1 (1×1) and P3 (√3×√3) symmetries is observed in the dispersion of the near-$E_F$ bands labelled 'feature A', comprised mainly of the strongly-hybridized Ni 3$d$ and O 2$p$ states. In the raw experimental data (a), as well as in the normalized data (b), it is evident that the valence-band maximum for this feature occurs at the Γ point, where the bands get closer to the Fermi level. This is also seen in Fig. 3a of the main text, where the maximum intensity at the binding energy of feature A is observed at the center of the BZ (Γ point) and not at the K points. This is consistent with the behavior of the corresponding bands for the P1 (1×1) symmetry shown in c. Conversely, corresponding feature A' for the P3 (√3×√3) symmetry exhibits the opposite behavior – the bands have their maxima at the K points and a characteristic 'dip' near the Γ point (see panel d above), which is inconsistent with the experiment.

Another major quantitative difference between the two symmetries arises from the size of the BZ, which is √3 smaller for the P3 (√3×√3) symmetry (due to a larger unit cell). This difference becomes apparent by comparing the horizontal $k_x$ scales in the figures above. A much better quantitative agreement between the experiment and the P1 symmetry is particularly obvious when comparing the width (along the $k_x$ scale) and the curvature of the parabolic band labeled 'feature D' at the valence-band bottom (see panels b and c). If P3 (√3×√3) was the case, the parabolic band would be noticeably 'compressed' and would be repeated several times within the experimental detector range (spanning several BZs) due to the smaller BZ size, which is inconsistent with the experiment.

The correct size and periodicity of the P1 (1×1) Brillouin zones in the extended BZ picture is evident in Fig. 3a of the main text, wherein the calculated schematic map of the extended BZ picture zone boundaries is overlayed on the experimental data, matching the hexagonal symmetry of the observed photoemission intensity distribution.

**Figure S6**

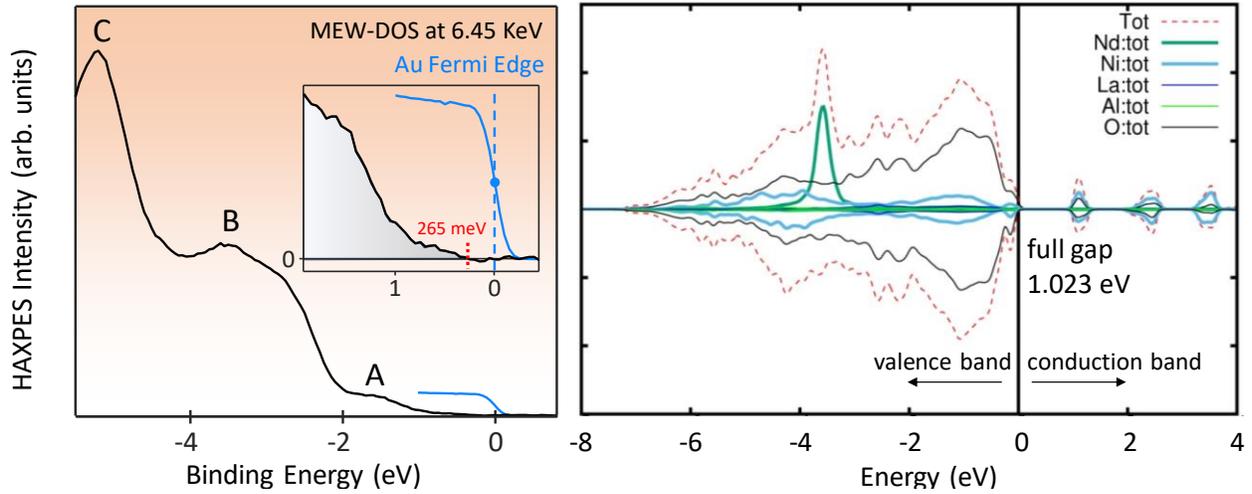

**Figure S6.** Comparison of the HAXPES measurement of the valence bands (left panel) and the theoretical DOS calculations (right panel), suggests that most of the 1.023 eV-wide full bandgap predicted by theory resides above the Fermi level. Measurement of the energy region below the Fermi level yields the value of the valence-band bandgap ($E_{\text{VB-max}} - E_F$) of 265 meV, qualitatively consistent with the theoretical picture shown in the right panel.

**Figure S7**

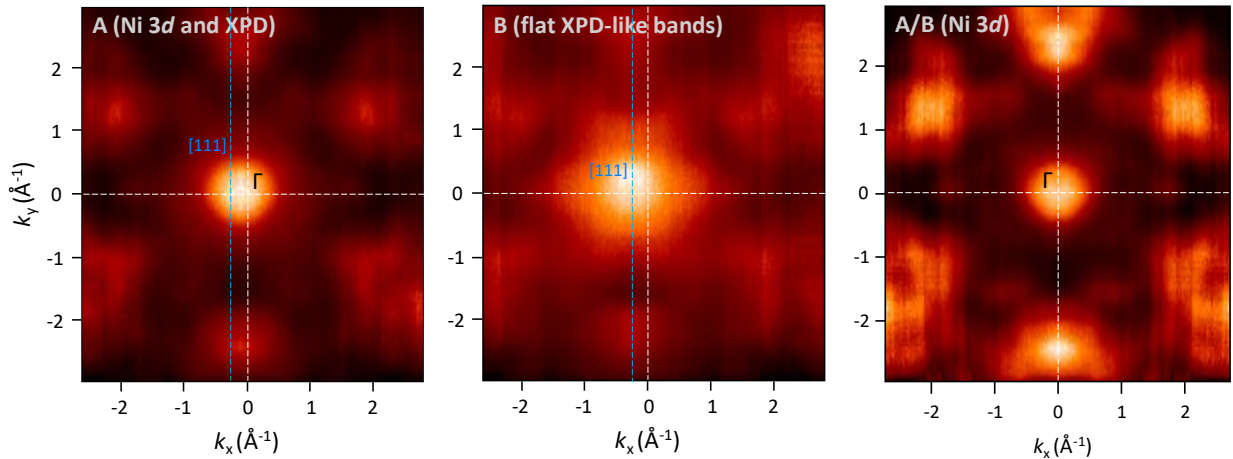

**Figure S7.** Separating band-dispersions in SX-ARPES from residual x-ray photoelectron diffraction (XPD). **a.** Non-normalized momentum-resolved photoemission intensity map of the VB feature A, containing combined contributions from the Ni $3d$ dispersive states and the XPD intensity modulations. The central intensity peak appears shifted toward the [111] emission direction due to a significant XPD contribution to the spectrum. **b.** Non-normalized photoemission intensity map of the VB feature B, comprised of flat XPD-like bands. The central intensity peak appears exactly along the [111] emission direction, confirming an overwhelming fraction of the XPD-derived intensity in the spectrum. **c.** XPD-corrected momentum-resolved photoemission intensity map of the VB feature A, obtained by scaled normalization A by B. Removal of the XPD effect shifts the central feature toward (0 Å$^{-1}$, 0 Å$^{-1}$), corresponding to the Γ-point in the first BZ.

**Figure S8**

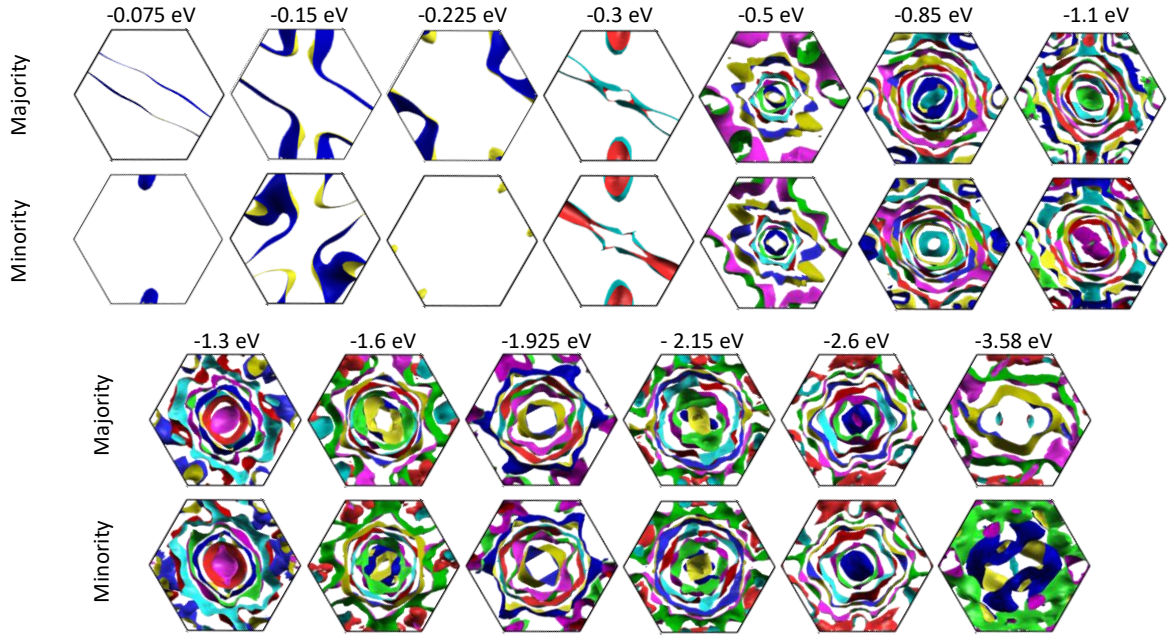

**Figure S8.** Isoenergetic cuts through the band structure in reciprocal space for the majority (top row) and minority (bottom row) bands, calculated for the binding-energy range from -0.075 eV to -3.58 eV, and spanning the major Ni 3*d* features in the VB manifold.

**Figure S9**

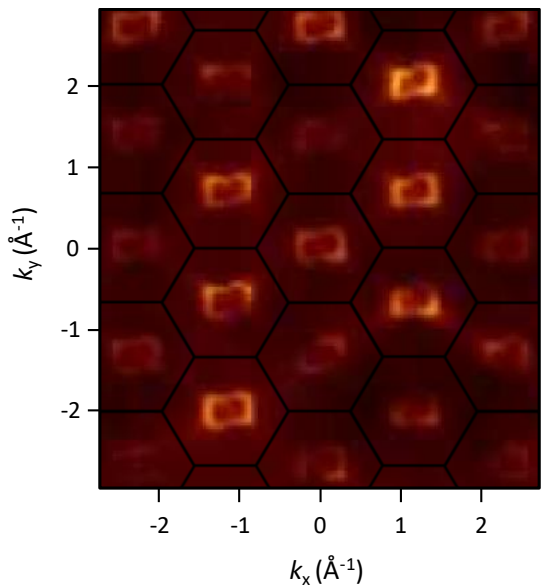

**Figure S9.** One step theory calculation of the momentum-resolved SX-ARPES spectra showing Ni 3*d* states near the Fermi level. Calculations were carried out using experimental SX-ARPES geometry for the photon energy of 642 eV. Clear modulations of the intensities between adjacent BZs due to the strong matrix-element effects are observed, consistent with the experimental observations.

**References for Supporting Materials**